# A proof of non-existence of self-imaging phenomenon in non-coherent case

## Jan A. Mamczur [1*]

[1] *Department of Physics, Rzeszow University of Technology, ul. W. Pola 2, 35-959 Rzeszow, Poland*
*Corresponding author: janand@prz.edu.pl*

**Abstract:** The existing description of non-coherent wave field propagation in terms of Fourier transformation has made possible to prove non-existence of the self-imaging phenomenon for non-coherent images.

## 1. Introduction

Self-imaging is meant in this paper as the phenomenon that occurs during propagation of an image in homogeneous isotropic stationary dielectric medium and consists in reconstruction of the original wave field intensity distribution in the plane distance *z* away from the original image. The authors approach to the self-imaging problem in the analogous way as W. D. Montgomery did in coherent case [1], i.e. by using the propagation operator in the diagonal form. In non-coherent case, the diagonalization by the Fourier transformation was presented in [2]. The essential results of [2] are collected in Section 2 of the present paper where monotonicity of the propagation operator has also been pointed out. These results has made possible to prove non-existence of self-imaging phenomenon for non-coherent images, which is showed in Section 3.

## 2. The diagonal operator of non-coherent wave field propagation.

Propagation of a non-coherent wave field can be described by the linear transformation of the wave field intensity distribution [4]:

$$I(x,y;z) = \hat{P}_z I(x,y;0) = \int\int_{-\infty}^{+\infty} g(x-x', y-y';z) I(x',y';0)\, dx'\, dy', \qquad (1)$$

where $I(x,y;0)$ and $I(x,y;z)$ are wave field intensity distributions in the original plane and in the plane distance $z$ away, respectively, and $\hat{P}_z$ is the non-coherent propagation operator. The integral operator kernel $g(x,y;z)$ in Eq. (1) is a well-known function of $x,y,z$ [2].

There exists a diagonal Fourier representation of the non-coherent propagation operator, converting the Fourier transform $J(\omega_x,\omega_y;0)$ of wave field intensity distribution in starting plane to intensity transform $J(\omega_x,\omega_y;z)$ of wave field formed at the distance $z$ [2]. This diagonal Fourier representation $G(\omega_x,\omega_y;z)$, defined by

$$J(\omega_x,\omega_y;z) = G(\omega_x,\omega_y;z)\,J(\omega_x,\omega_y;0), \tag{2}$$

is the Fourier transform of the kernel $g(x,y;z)$:

$$G(\rho;z) = F[g(x,y;z)] = \frac{\kappa k^2}{8\pi^2} z\rho K_1(z\rho) + \frac{\kappa}{32\pi^2}\rho^2 K_2(z\rho). \tag{3}$$

where $\omega_x, \omega_y$ are spatial angular frequencies having the sense of wave-vector projection on the axes $x$ and $y$, $K_1$ and $K_2$ are the modified Bessel functions of second kind (MacDonald functions) of first and second order, respectively, $\kappa$ is a positive real constant, $k$ is the wave number, and $\rho$ is the radius in spatial angular frequency domain defined by

$$\rho = \sqrt{\omega_x^2 + \omega_y^2}. \tag{4}$$

Selection rule of the constant $\kappa$ was showed in [2]. Equation (2) together with the transform $G(\rho;z)$ is a more convenient tool for calculations than Eq. (1) with kernel $g(x,y;z)$ thanks to the diagonalization of the non-coherent propagation operator and to fast Fourier transformation efficiency.

When propagation distance $z$ is fixed, the transform $G(\rho;z)$ is a decreasing function of angular frequency radius $\rho$. It can be proved by using the formula for modified Bessel function differentiation [7]:

$$K'_\nu(x) = -\tfrac{1}{2}\left(K_{\nu-1}(x) + K_{\nu+1}(x)\right), \tag{5}$$

and the formula for replacing Bessel function of higher order with functions of lower orders [7]

$$K_{\nu+1}(x) = K_{\nu-1}(x) + \frac{2\nu}{x} K_\nu(x), \tag{6}$$

which yields after substituting

$$K'_\nu(x) = -K_{\nu-1}(x) - \frac{\nu}{x} K_\nu(x). \tag{7}$$

The function $G(\rho;z)$ derivative is equal

$$\frac{\partial G(\rho;z)}{\partial \rho} = \frac{\kappa k^2}{8\pi^2} z K_1(z\rho) - \frac{\kappa k^2}{8\pi^2} z^2 \rho\left(K_0(z\rho) + \frac{1}{z\rho} K_1(z\rho)\right) +$$

$$+ \frac{\kappa}{16\pi^2}\rho K_2(z\rho) - \frac{\kappa}{32\pi^2} z\rho^2\left(K_1(z\rho) + \frac{2}{z\rho} K_2(z\rho)\right) =$$

$$= -\frac{\kappa k^2}{8\pi^2} z^2 \rho K_0(z\rho) - \frac{\kappa}{32\pi^2} z\rho^2 K_1(z\rho) < 0 \tag{8}$$

There are the minus signs in front of both the derivative components, which together with the fact that modified Bessel functions are positive in real domain yields function $G(\rho;z)$ monotonicity.

### 3. The problem of self-imaging of non-coherent wave fields

If the self-imaging effect occurs in non-coherent case than there is at least one wave field intensity distribution that maps to the identical distribution as a result of propagation at the distance $z$. Using the propagation equation (2) and allowing the two intensity distributions to differ by a multiplicative real constant, we can write this assumption as an eigenequation in the Fourier representation:

$$J(\omega_x, \omega_y; z) = G(\omega_x, \omega_y; z) J(\omega_x, \omega_y; 0) = C\, J(\omega_x, \omega_y; 0) \;, \tag{9}$$

where $C$ is a real constant. Like W. D. Montgomery [1] in coherent case, we can formulate a condition for the non-coherent self-imaging wave field on the basis of the above equation: the Fourier transform $J(\omega_x, \omega_y; 0)$ of intensity distribution of such wave field, being an eigenfunction of Eq. (9), must take on non-zero values only in the angular spatial frequency region $\{(\omega_x, \omega_y)\}$ that satisfy the condition

$$G(\omega_x, \omega_y; z) \equiv G(\rho; z) = C \;. \tag{10}$$

As it has been shown in Section 2, the function $G(\rho;z)$ is monotonic in the whole frequency domain. Therefore the eigenequation (9) may only have such nontrivial eigenfunctions $J(\omega_x, \omega_y; 0)$ that take on non-zero values only at one spatial frequency radius $\rho$, i.e. in one circle-shaped spatial frequency region with a radius $\rho_o \neq 0$. The transform $J(\omega_x, \omega_y; 0)$ is meant here to be trivial if it takes on non-zero value at $\rho=0$ only, i.e. the corresponding intensity distribution $I(x,y;0)$ is uniform. On the other hand, intensity distribution of every image is non-negative and has positive average, and hence its Fourier transform is positive at the zero spatial frequency $\rho$. Because there is only one $\rho_o$, it yields $\rho_o=0$, i.e. the wave field is trivial. Therefore there is a contradiction in the demand that a nontrivial transform of physical wave field intensity distribution is an eigenfunction of eigenequation (9). In other words, the self-imaging effect does not exist in non-coherent case.

### 4. Conclusions

The diagonalized operator of non-coherent propagation has made possible to prove non-existence of non-coherent self-imaging phenomenon for non-trivial images. Only infinite non-coherent image with uniform intensity distribution does not change as a result of propagation at a certain distance.

**Acknowledgements**

This work is supported by the State core funding for statutory R & D activities of the Department of Physics, Rzeszow University of Technology, channeled entirely through the Polish Ministry of Science and Higher Education